# Scalable boson sampling with a single-photon device

Yu He[1,2], Zu-En Su[1,2], He-Liang Huang[1,2], Xing Ding[1,2], Jian Qin[1,2], Can Wang[1,2], S. Unsleber[3], Chao Chen[1,2], Hui Wang[1,2], Yu-Ming He[1,3], Xi-Lin Wang[1,2], Christian Schneider[3], Martin Kamp[3], Sven Höfling[1,3,4], Chao-Yang Lu[1,2], and Jian-Wei Pan[1,2]

[1] Hefei National Laboratory for Physical Sciences at Microscale and Department of Modern Physics, University of Science and Technology of China, Hefei, Anhui, 230026, China

[2] CAS-Alibaba Quantum Computing Laboratory, CAS Centre for Excellence in Quantum Information and Quantum Physics, University of Science and Technology of China, China

[3] Technische Physik, Physikalisches Instität and Wilhelm Conrad Röntgen-Center for Complex Material Systems, Universitat Würzburg, Am Hubland, D-97074 Wüzburg, Germany

[4] SUPA, School of Physics and Astronomy, University of St. Andrews, St. Andrews KY16 9SS, United Kingdom

## Abstract

**Boson sampling is a problem intractable for classical computers, but can be naturally solved on a specialized photonic quantum simulator which requires less resources than building a universal quantum computer. The biggest challenge to implement boson sampling with a large number of photons has been the lack of reliable single-photon sources. Here we demonstrate a scalable architecture of boson sampling using a solid-state single-photon source with simultaneously high efficiency, purity, and indistinguishability. The single photons are time-bin encoded and interfered in an electrically programmable loop-based network. We implement and validate boson sampling with input three and four single photons, and track the dynamical multi-photon evolution inside the circuit. With further refinement of the system efficiency, our approach may be feasible to be scaled up to $\gtrsim$ 20-boson sampling to outperform classical computers, and thus provide experimental evidence against the Extended Church-Turing Thesis.**

Quantum computers are expected to solve specific problems such as factoring integers using Shor's algorithm (*1*) exponentially faster than classical machines. Despite substantial progress in the past decades (*2-5*), building a universal quantum computer, which requires fault-tolerant manipulation of a large number of quantum bits (qubits), remains a formidable challenge. The number of qubits needed by Shor's algorithm to outperform the classical computers is orders of magnitude larger than the ~10 qubits that can be experimentally controlled today. This motivates considerable interest in non-universal quantum computers which demand less physical resources but can demonstrate quantum speedup for specific problems.

Boson sampling (*6*), proposed by Aaronson and Arkhipov in 2011, can be realized by sending *n* indistinguishable single photons through a passive *m*-mode linear optical network, which doesn't require adaptive measurement, deterministic entangling gate, and makes less stringent demands on device performance than universal linear optical quantum computation (*7*). Boson sampling is strongly believed to be hard for classical computers, but can be naturally and efficiently obtained from the output distribution of a multimode bosonic interferometer. With input $\gtrsim 20$ indistinguishable single photons, boson sampling would already reach a computational complexity that can challenge classical computers (*6,8*). This would constitute a strong evidence against a foundational tenet in computer science: the Extended Church-Turing Thesis, which postulates that all realistic physical systems can be efficiently simulated with a (classical) probabilistic Turing machine.

However, scaling up boson sampling to a large number of photonic qubits remains a nontrivial experimental challenge, most importantly due to the lack of reliable single-photon sources. So far, all implementations (*9-14*) employed inefficient pseudo-single photons from spontaneous parametric down-conversion (SPDC) (*15*), which was probabilistically generated and inevitably admixed with multi-photon contributions. The down-conversion probability (*p*) was kept small, typically of a few percent, in order to suppress the multi-photon emission that scales as $\sim p^2$, the dominant source of noise in the previous multiphoton experiments (*2*). Thus, the proof-of-principle experiments for boson sampling with SPDC have tested up to three photons for arbitrary input configurations. With heralded single photons from the state-of-the-art SPDC source (*16*) and typical optical circuit efficiency (*9-14*), one would expect only about two four-photon detection rates per day.

To overcome this obstacle, here we use a deterministic single-photon source and an electrically programmable multi-photon interferometer to implement the boson sampling in a scalable architecture. The single photons are produced from a single self-assembled InAs/GaAs quantum dot (*17-19*) embedded inside a micropillar (see Fig.1A and supplementary materials, section S1). For boson sampling, it is crucial that the single photons simultaneously possesses high purity, indistinguishability, and efficiency. Using pulsed *s*-shell resonant pumping (*20*), the three key features can be compatibly combined (*21,22*). At π pulse excitation using a picosecond laser with a repetition rate of 76.4 MHz, the quantum dot-micropillar emits ~10.6 million pulsed resonance fluorescence single photons at the output of a single-mode fiber (with an absolute brightness of ~13.9%), of which 3.5 million is eventually detected on a silicon single-photon detector. This is ~22 times brighter than the heralded polarized single photons from the SPDC (*17*).

An ideal single-photon Fock state should have no multi-photon admixture and thus exhibit perfect photon antibunching, i.e., $g^2(0)=0$, where $g^2(0)$ is the second-order correlation at zero time delay. We observe an almost vanishing $g^2(0)$ of 0.011(1) after passing the photons through a 5-GHz etalon, proving the high purity of the single-photon source (supplementary materials, section S2). The non-classical Hong-Ou-Mandel interference (*23*) in the boson sampling multi-photon interferometry relies on a high degree of indistinguishability between the photons (*24-26*). *S*-shell resonant excitation is used to eliminate dephasings and emission time jitter, and generate long streams of single photons with near-unity indistinguishability (*27,28*). For two photons with their emission time separated by 13 ns and 26 ns, their mutual indistinguishabilities can reach 0.978(4) and 0.970(3), respectively.

To implement the multi-photon interferometer for boson sampling, we utilize a scalable time-bin encoding scheme (*29-31*), as illustrated in Fig.1A. For each experimental period, *M* time bins (each loaded with one or zero photon) are injected into a loop by an acousto-optical mudulator (AOM-in) and circulated for *N* loops. Such a loop-based architecture is equivalent to an *M*-mode beam splitter network with a depth of *N*, as expanded in Fig.1B. Here, the polarization degree of freedom acts as the spatial mode in the conventional boson sampling model. The beam splitter operations (denoted by the circles in Fig.1B) are effectively realized using a polarization-rotation electro-optic modulator (*p*-EOM) with dynamically programmable coupling ratio. After the *p*-EOM, a polarization-dependent asymmetric Mach-Zehnder interferometer delays the vertical polarization for one time-bin length (~13 ns), respective to the horizontal polarization, which realizes the displacement operation of time bins. The optical transmission efficiency of one loop is measured to be 83.4%.

After *N* loops of evolution, the *M* bins are ejected out of the loop by an AOM, and the output distribution are obtained by registering all the single-photon detection events in real time and postprocessing. The time-bin encoding scheme naturally complements the single-photon pulse train and doesn't require active demultiplexing. We emphasize that such a loop-based architecture is intrinsically stable, electrically rapid programmable, and resouce efficient (supplementary materials, section S3). Further, it relaxes the need for overcoming the inhomogenity of independent self-assembled quantum dots to build many identical sources.

We implement boson sampling with *n*=3 and *n*=4 single photons, propagating them through loop-based interferometers with *m*=6 and *m*=8 modes, respectively. Figure 2A and 2B show two typical equivalent boson sampling circuits, programmed by electric pulse sequences shown in the upper panels that drive the *p*-EOM. The linear unitary operators of the photonic networks are chosen to be fully connected and their permanents (*6*) cannot be calculated efficiently. With a classical computer, the probability of an output bosonic scattering event is calculated from the permanent of corresponding submatrix—a computationally hard problem with runtime increasing exponentially with the number of input bosons.

We measure the coincidence count rates of 36 and 180 different combinations of output distributions for the 3- and 4-boson sampling, respectively, including both no-collision (one photon per output-mode) and collision events (multiple photons per output-mode). A total of 2015 and 1219 events are recorded in the 3- and 4-boson sampling experiment within 1 and 78 hours, respectively. Compared to previous implementation with SPDC (*9-14*), the high-efficiency single-photon source enables us to complete the 3-boson sampling ~200 times faster, and achieve the first 4-boson sampling with single-photon Fock state which was a formidable challenge before. The data (solid bar) are plotted in Fig.2C-D together with calculated distributions (empty bars). We quantify the match between the two sets of normalized distributions obtained experimentally ($q_i$) and theoretically ($p_i$), using the measure of the fidelity: $F = \sum_i \sqrt{p_i q_i}$. From the experimental data shown in Fig.2C-D, we obtain a fidelity of 0.991(2) for the 3-boson sampling, and 0.953(4) for the 4-boson sampling, respectively.

Unlike prime factoring—among other problems in the NP complexity class for which the solution can be efficiently verified—the boson sampling is related to calculating the matrix permanent, a harder problem in the #P-complete complexity class that a full certification of the outcome itself for large number of bosons becomes exponentially intractable for classical computation. Fortunately, there have been proposals (*32-34*) and demonstrations (*35,36*) for scalable validation of boson sampling that can provide supporting or circumstantial evidence for the correct operation of the protocol. Firstly, we apply the experimental data to the Aaronson and Arkhipov test (*32*), designed to distinguish the outcome of fixed-input boson sampling from a uniform distribution. The uniform distribution can be conclusively ruled out with ~50 events (see Fig.3A-B). Secondly, we employ the Bayesian analysis (*33*) to exclude the possibilities of standard normal distribution (see Fig.3C-D) and uniform distribution (supplementary materials, section 4). With only ~10 events, we can reach a confidence level of 99.8% that the output distribution data are from a genuine boson sampler. Finally, we adapt the standard likelihood ratio test (*34,36*) to rule out the hypothesis that the data could be reproduced with a sampler with distinguishable bosons. Figure 3E-F show an increasing difference between organge (indistiguishable bosons) and blue points (distinguishable bosons) as the sampling events increase, and confirm that our data are indeed expected from the highly indistinguishable single photons.

The flexible loop-based architecture in this experiment further allow us to track the dynamical multi-photon evolution in the circuit at intermediate time. Controlled by the ejection time of the AOM-out (Fig.1), the output distribution can be measured and monitored, on a loop-by-loop basis. The evolution of the multi-photon scattering of a new 3-boson sampling circuit (supplementary materials, section S3.2) at the end of the 1-5 loop are shown in Fig.4A-E, respectively. The measured fidelities (upper panels) from the 1st to the 5th loop are 0.960(3), 0.982(3), 0.983(3), 0.990(3), and 0.989(3), respectively, in a good agreement with the theoretical calculations (lower panels). Our experiment opens a new way to study multi-particle high-dimensional quantum walks with single quantum emitters (*31,37,38*).

The overall efficiency of the current experiment is mainly limited by the system efficiency of the single-photon source (~13.9%, including photon extraction, cross polarization, optical path transmission, and fiber coupling), interferometric network efficiency (~83.4% for a single loop), and single-photon detection efficiency (~33%). With on-going technical advances on deterministic quantum dot-micropillar (*22,39*) and background-free side excitation (to avoid cross-polarization), it appears realistic to reach a modest system efficiency of the single-photon source to be ~60%. With this, and by further combining superconducting nanowire single-photon detection with reported ~95% efficiency (*40*) and antireflection coatings of the optical elements, we estimate that 20-boson sampling with a reasonable (~100/hour) coincidence rate is feasible in future experiments. We believe that our first implementation of multi-photon boson sampling based on a solid-state single-photon source with a scalable time-bin architecture brings boson sampling closer to an experimental regime approaching quantum supremacy.

During manuscipt preparation, we became aware of a related work (*41*) on 3-boson sampling using non-resonantly pumped, passively demultiplexed single photons from quantum dot.

**Acknowledge**: We gratefully thank Peter Rohde and Christine Silberhorn for helpful comments. This work was supported by the National Natural Science Foundation of China, the Chinese Academy of Sciences, the National Fundamental Research Program, and the State of Bavaria.

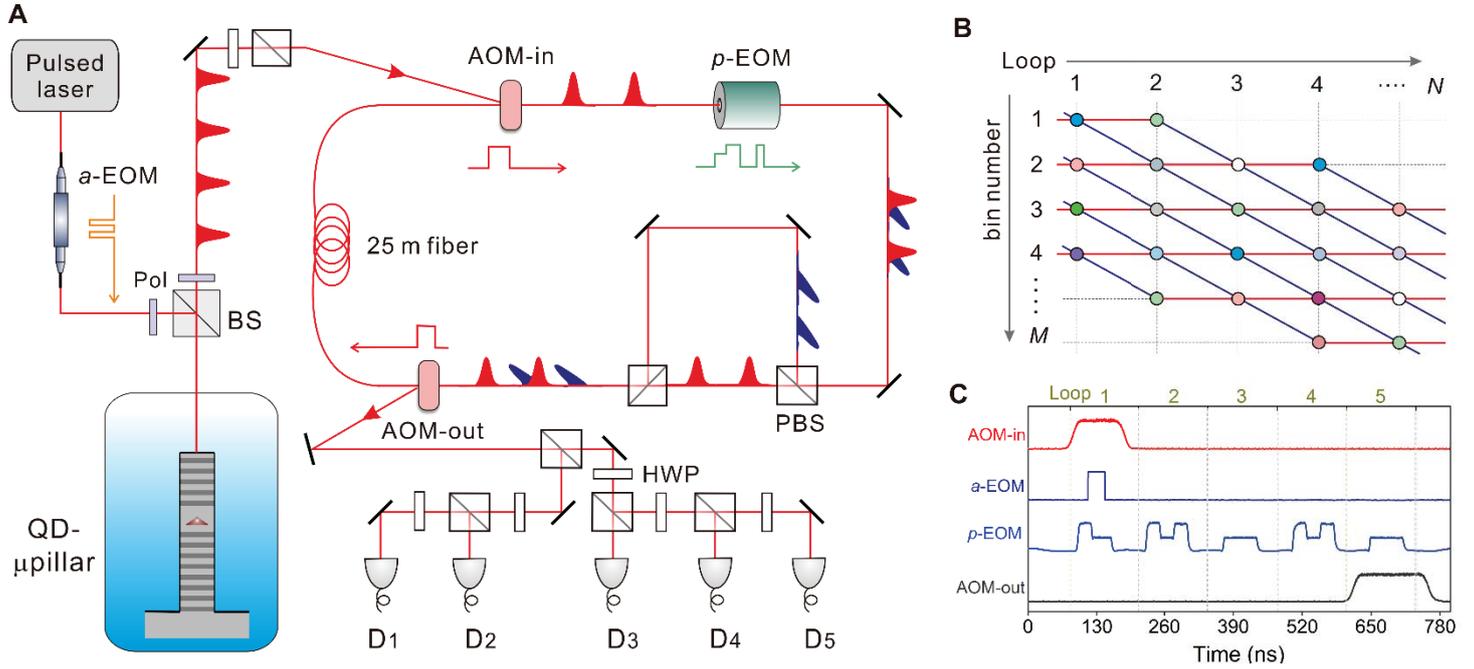

Fig.1: Boson sampling implementation with a single-photon device. **A.** Experimental arrangement. The pumping laser is chopped by a waveguide-based amplitude electro-optic modulator (*a*-EOM) to prepare a single-photon pulse train in designed time bins. The quantum dot is sandwiched between 25.5 lower and 15 upper λ/4-thick AlAs/GaAs mirror pairs that form the distributed Bragg reflectors, and embedded inside a 2.5 μm diameter micropillar cavity. The device is cooled to 7 K where the quantum dot emission is resonant with the micropillar cavity mode. A confocal microscope is operated in a cross-polarization configuration to extinguish laser leakage. The prepared three or four single photons are injected into a loop by an acousto-optic modulator (AOM). An electro-optical modulator (*p*-EOM) rotates the polarization controlled by a pulse sequence. The red (blue) pulse in the loop denotes horizontal (vertical) polarization. A 25 m single-mode fiber is inserted in the loop lengthening it to 130 ns (10 bins). After several loops of evolution (see text for details), the photons are ejected out of the loop by the AOM-out for coincidence measurements. **B.** An equivalent beam splitter network unravelling the dynamics of *M* time bins circulating for *N* loops. The circles denote beam splitter operations and their color coding represents arbitrary, electrically programmable coupling ratios. The red and blue line evolution represents the trajectory for horizontal and vertical polarization, respectively. **C.** Electrical pulse sequences for implementing Boson sampling. The whole system is time synchronized to the pulsed laser.

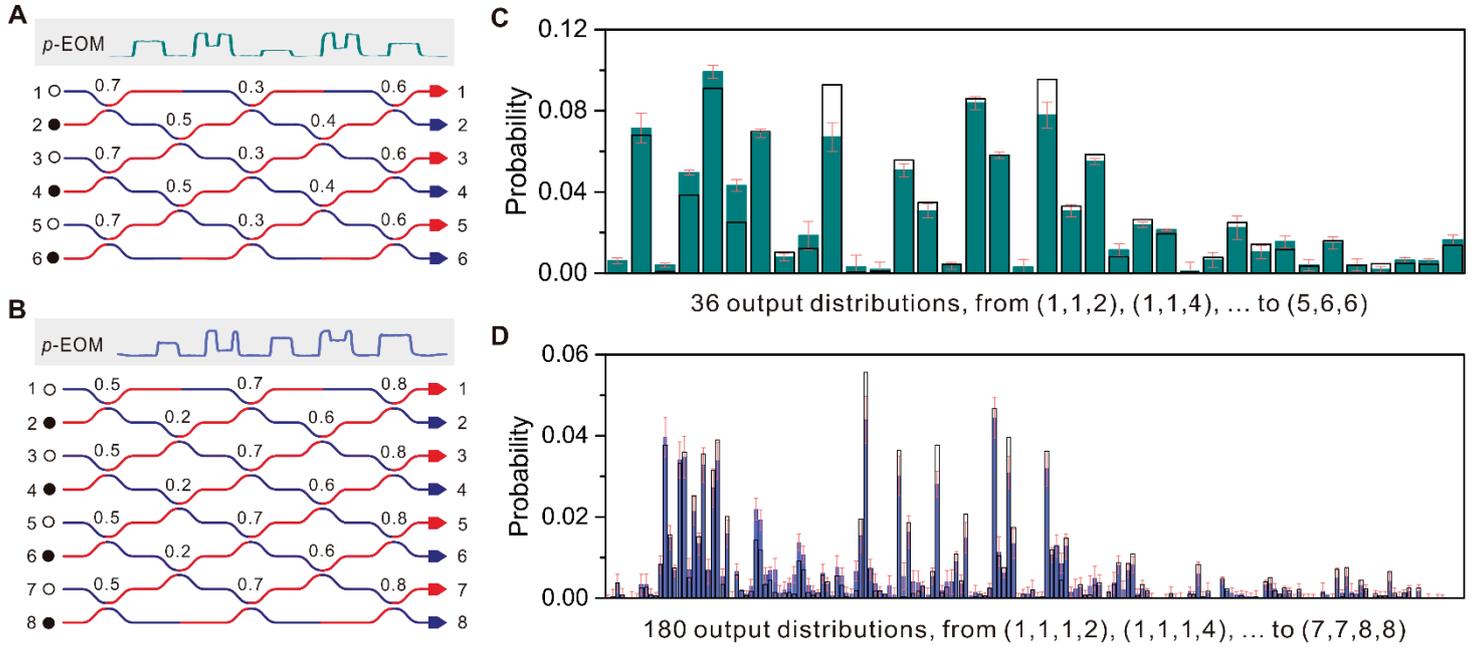

Fig.2: Experimental results for 3- and 4-boson sampling. **A, B** The equivalent 3- and 4-boson sampling circuits implemented. The upper panel shows the electrical pulse sequence that drives the *p*-EOM and programs the circuits. The inputs to the circuits are one (zero) photon Fock state, represented by solid (empty) circles. **C, D** The measured relative frequencies of various output combinations, denoted by (*i, j, k*) where *i*, *j*, and *k* are the output modes as labelled in A and B. Note that collision events—multiple photons per output-mode, *e.g.* (1,1,2)—are also registered. The solid bars are the normalized coincidence rate of different output distribution. The empty bars are theoretical calculations in the ideal case. The error bar is one standard deviation from Poissionian counting statistics.

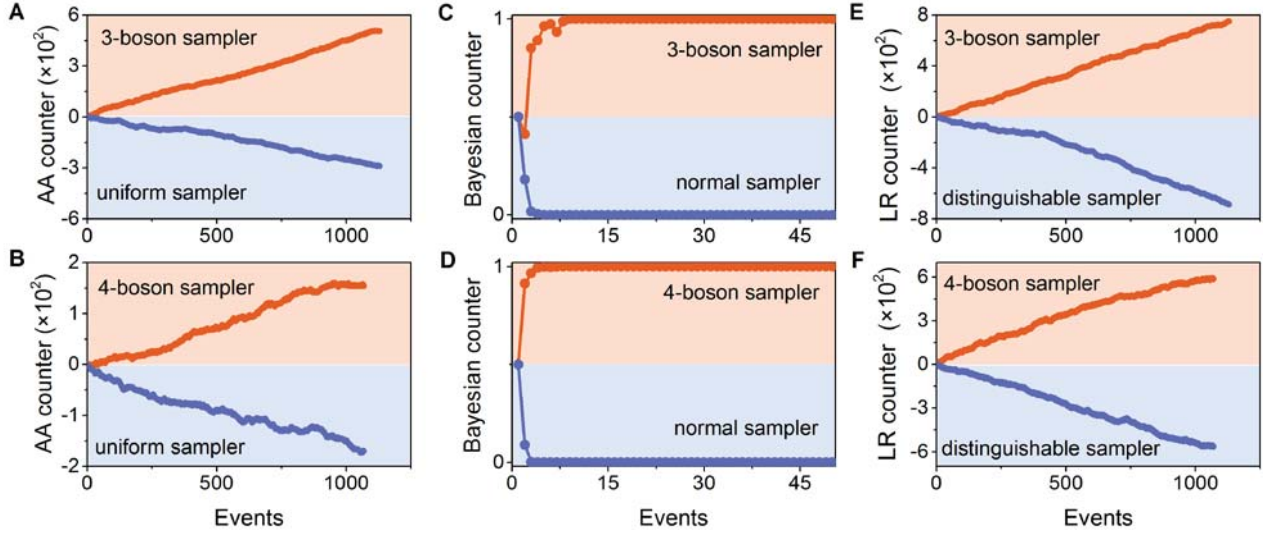

Fig.3: Validating boson sampling results. Blue points are tests applied on simulated data generated from the three alternative hypotheses. Orange points are tests applied on the experimental data. A counter is updated for every event and a positive value validates the data being obtained from a genuine boson sampler. **A, C, E** are results for 3 bosons and **B, D, F** are for 4 bosons. **A, B** Using the Aaronson and Arkhipov (AA) test to rule out the uniform distribution. With ~50 events, this method reaches a 100% average success rate. **C, D** Application of the Bayesian analysis to test against standard normal distribution. **E, F** Discrimination of the data from a distinguishable sampler using standard likelihood ratio test.

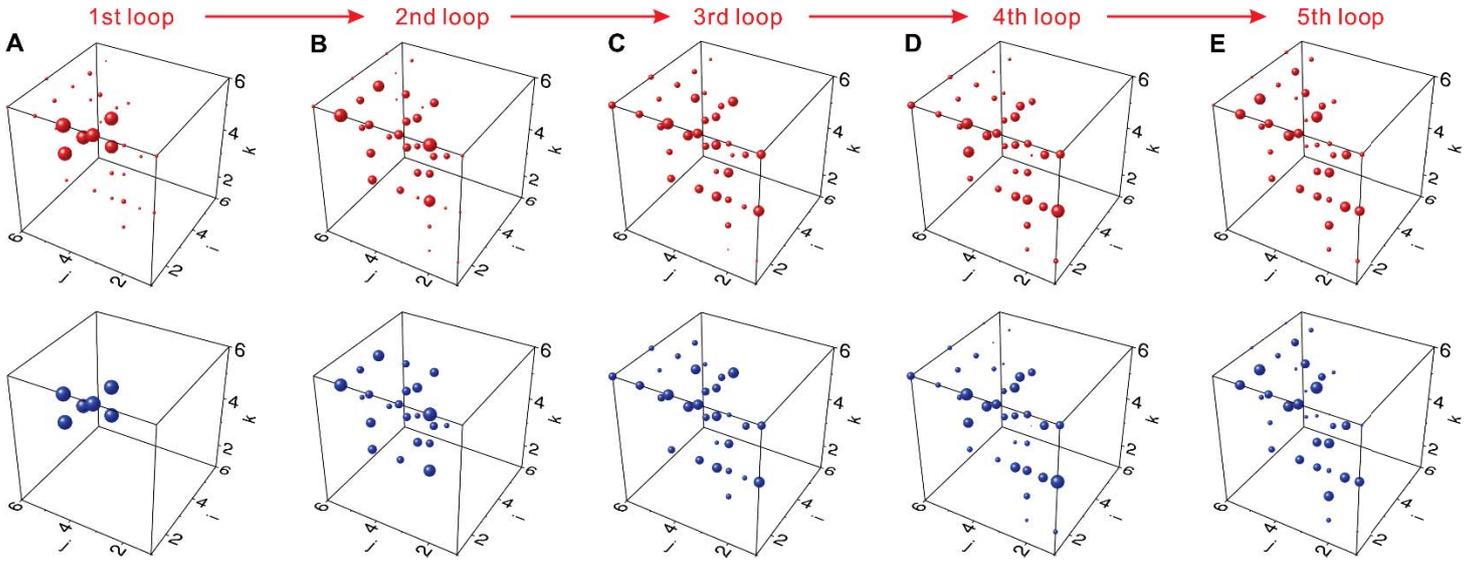

Fig.4: Tracking boson sampling dynamics. The output distribution is measured after 1 (**A**), 2 (**B**), 3 (**C**), 4 (**D**), and 5 (**E**) loops evolution in the 3-boson sampling circuit shown in Fig.2A. The probability of finding three photons in the output-mode distributions ($i, j, k$) (see Fig.2 caption) are plotted using a sphere centered at coordinates ($i, j, k$), where the volume of the sphere is proportional to the occurring frequency. The measured fidelities from the 1st to the 5th loop are 0.960(3), 0.982(3), 0.983(3), 0.990(3), and 0.989(3), respectively. The upper and lower panels are the experimental data and theoretical calculation, respectively.